\documentclass[aps,pre,twocolumn]{revtex4-1}
\usepackage{colortbl} 
\usepackage{graphicx}
\usepackage{dcolumn}
\usepackage{amsmath,amssymb}
\usepackage{bm}

\def\figp#1#2#3#4{
\begin{figure}[!tp]
\begin{center}
\includegraphics[width=#3\textwidth,bb=#4]{./figures/#1.pdf}
\caption{#2}
\label{fig:#1}
\end{center}
\end{figure}}

\begin{document}


\title{Fluctuation-dissipation theory of input-output interindustrial relations}


\author{Hiroshi Iyetomi}
\email[]{hiyetomi@sc.niigata-u.ac.jp}
\author{Yasuhiro Nakayama
}
\affiliation{Department of Physics, Niigata University, Ikarashi, Niigata 950-2181, Japan}
\author{Hideaki Aoyama}
\affiliation{Department of Physics, Kyoto University, Kyoto 606-8501, Japan}
\author{Yoshi Fujiwara}
\affiliation{ATR Laboratories, Kyoto 619-0288, Japan}
\author{Yuichi Ikeda}
\affiliation{Hitachi Research Laboratory, Hitachi Ltd., Ibaraki 319-1221, Japan}
\author{Wataru Souma}
\affiliation{College of Science and Technology, Nihon University, Chiba, 274-8501, Japan}


\date{\today}

\begin{abstract}
In this study, the fluctuation-dissipation theory is invoked to shed light on input-output interindustrial relations at a macroscopic level by its application to IIP (indices of industrial production) data for Japan. Statistical noise arising from finiteness of the time series data is carefully removed by making use of the random matrix theory in an eigenvalue analysis of the correlation matrix; as a result, two dominant eigenmodes are detected. Our previous study successfully used these two modes to demonstrate the existence of intrinsic business cycles. Here a correlation matrix constructed from the two modes describes genuine interindustrial correlations in a statistically meaningful way. Further it enables us to quantitatively discuss the relationship between shipments of final demand goods and production of intermediate goods in a linear response framework. We also investigate distinctive external stimuli for the Japanese economy exerted by the current global economic crisis. These stimuli are derived from residuals of moving average fluctuations of the IIP remaining after subtracting the long-period components arising from inherent business cycles. The observation reveals that the fluctuation-dissipation theory is applicable to an economic system that is supposed to be far from physical equilibrium.

\end{abstract}


\maketitle

\section{Introduction}\label{sec:intro}
Both the business cycle and the interindustrial relationship are long-standing basic issues in the field of macroeconomics, and they have been addressed by a number of economists. 
Recently, we analyzed~\citep{bc} business cycles in Japan using indices of industrial production (IIP), an economic indicator that measures current conditions of production activities throughout the nation on a monthly basis. Careful noise elimination enabled us to extract business cycles with periods of 40 and 60 months that were hidden behind complicated stochastic behaviors of the indices. 

In this accompanying paper, we focus our attention on interindustrial relationship by analyzing IIP data in a framework of the linear response theory; the fluctuation-dissipation theory plays a vital role in the analysis of IIP data. We also discuss the difference between moving average fluctuations in the original data and long-period components arising from inherent business cycles. The residuals may be interpretable as a sign of external stimuli to the economic system. The recent worldwide recession offers us a good opportunity to conduct this study, because it delivered an unprecedented shock to the economic system of Japan. 

The interindustrial relations of an economy are conventionally represented by a matrix in which each column lists the monetary value of an industry's inputs and each row lists the value of the industry's outputs, including final demand for consumption. Such a matrix, called the input-output table, was developed by Leontief~\citep{Leontief1936,Leontief1986}. This table thus measures how many goods of one industrial sector are used as inputs for production of goods by other industrial sectors and also the extent to which internal production activities are influenced by change in final demand. Leontief's input-output analysis can be regarded as a simplified model of Walras's general equilibrium theory~\citep{Walras1954} to implement real economic data for carrying out an empirical analysis of such economic interactions. 
Currently, the basic input-output table is constructed every 5 years according to the System of National Accounts (SNA) by the Ministry of Internal Affairs and Communications in Japan.

It should be noted that the input-output table describes yearly averaged interindustrial relations. 
Although such a poor time resolution of the table may be tolerable for budgeting of the government on an annual basis, 
various day-to-day issues faced by practitioners require them to react promptly. We thus need a more elaborate methodology that enables investigation of the input-output interindustrial relationship with a much higher time resolution.


Econophysics~\citep{MS2000, BP2003, AY2007, AFIIS2010} is a newly emerging discipline in which physical ideas and methodologies are applied for understanding a wide variety of complex phenomena in economics. We adopt this approach to address the above-mentioned issues in macroeconomics. That is, we pay maximum attention to real data while drawing any conclusions. Of course it is important to remember that real data are possibly contaminated with various kinds of noise. The random matrix theory (RMT), combined with principal component analysis, has been used successfully to extract genuine correlations between different stocks hidden behind complicated noisy market behavior~\citep{laloux1999ndf, plerou1999uan, plerou2002rma, utsugi2004rmt, kim2005sag, KD2007, PhysRevE.76.046116, SZ2009}. Recently, dynamical correlations in time series data of stock prices have been analyzed by combining Fourier analysis with the RMT~\citep{nakayamai2009rmt}. 

In this study, we further develop the noise elimination method initiated in previous studies. The null hypothesis that has been adopted thus far for extracting true mutual correlations corresponds to shuffling time series data in a completely random manner. Although we should distinguish between mutual correlations and autocorrelations, both these correlations are destroyed at the same time by the completely random shuffling. To solve this problem, rotational random shuffling of data in the time direction is introduced as an alternative null hypothesis; such randomization preserves autocorrelations involved in the original data. The new null hypothesis thus elucidates the concept of noise elimination for mutual correlations.

Further, we borrow the concept of the fluctuation-dissipation theory~\citep{Landau1980} from physics to elucidate the interindustrial relationship and the response of an economic system to external stimuli. The theory establishes a direct relationship between the fluctuation properties of a system in equilibrium and its linear response properties. We assume that the validity of the fluctuation-dissipation theory in physical systems is also true for such an exotic system as described by the IIP. Very recently, dynamics of the macroeconomy has been studied in the linear response theory by taking an explicit account of heterogeneity of microeconomic agents~\citep{HA2008}.

The present paper is organized as follows. In Sec.~\ref{sec:rmt-iip}, we first provide a brief review of the noise elimination from the IIP using the RMT. A new null hypothesis based on rotational random shuffling is introduced in Sec.~\ref{sec:rs}. Section~\ref{sec:gcm} presents construction of a genuine correlation matrix for the IIP by consideration of only those dominant modes that are approved to be statistically meaningful by the RMT. We present development of a fluctuation-dissipation theory for input-output interindustrial relations in Sec.~\ref{sec:fdt}. Then, in Sec.~\ref{sec:ic}, we quantitatively discuss relationship between shipments of final demand goods and production of intermediate goods. In Sec.~\ref{sec:iir}, we elucidate response of the industrial activities to external stimuli by subtracting long-period components arising from inherent business cycles from moving average fluctuations in the original data. Section~\ref{sec:concl} concludes this paper.

\section{Application of Random Matrix Theory to IIP}\label{sec:rmt-iip}

In Japan, the IIP are announced monthly by the Ministry of Economy, Trade and Industry~\citep{iip}. For this study, we will choose seasonally adjusted data instead of original data. 
Two classification schemes of the IIP are available: indices classified
{\it by industry\/} and indices classified {\it by use of goods\/}.
We adopt the latter classification scheme because we are interested in input-output interindustrial relations here, which are measured by correlations between shipments of final demand goods and production of intermediate goods in the IIP data. The concept is illustrated in Fig.~\ref{fig:IIP_description}. We emphasize that the inner loop of production existing in the economic system may give rise to a nonlinear feedback mechanism to complicate the dynamics of the system; outputs are reused by the system as inputs for its production activities. Table~\ref{tab:21} lists the categories~\footnote{See reference \citep{bc} for details on the classification.} of goods along with weights assigned to each of them for computing the average IIP. These weights are proportional to value added produced in the corresponding categories, and their total sum amounts to 10,000. Unfortunately, the resolution of the IIP data for the producer goods is quite poor, which are just categorized as Mining \& Manufacturing and Others. 

\figp{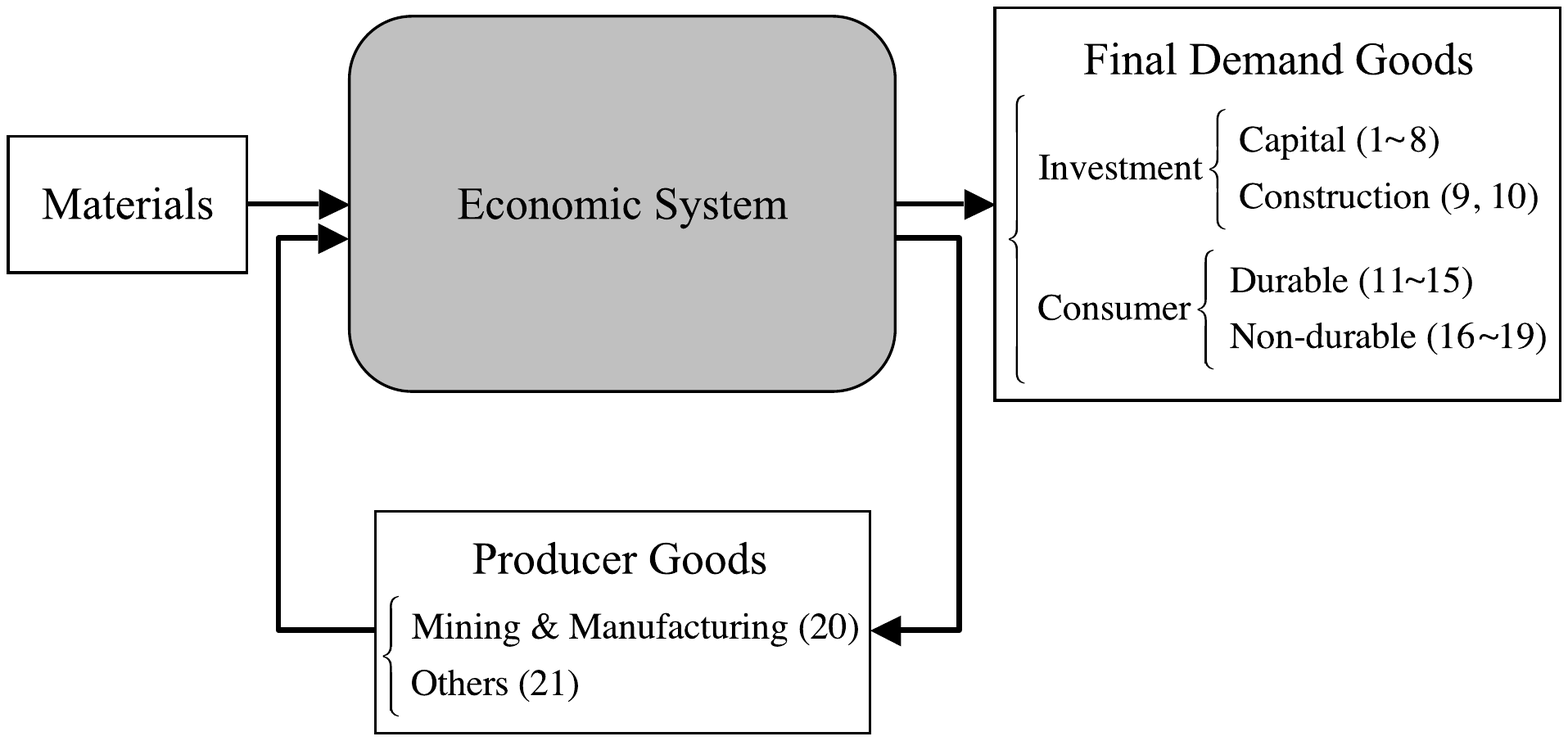}{Input-output relationship in industrial activities of economic system as measured by IIP in Japan. The numbers in the parentheses denote the classification index $g$ in Table~\ref{tab:21}.}{0.45}
{162.52 192.0 713.08 448.08}

\setlength{\doublerulesep}{4pt} 
\renewcommand{\arraystretch}{1.1}
\begin{table*}\small
\centering
\caption{Classification of goods according to IIP. First, the goods are classified into two categories, ``Final Demand" and ``Producer," and then those two categories are divided into 19 and 2 subcategories, respectively. The central column 
specifies the index $g$ for the subcategories, which has totally 21 values. The number in the parenthesis associated with each species of goods shows its weight used to compute the averaged IIP; the weights are normalized so that their total sum is 10,000.}
\label{tab:21}
\begin{tabular}{|c|c|l|}
\hline
\multicolumn{3}{|>{\columncolor[gray]{0.6}}c|}{{\sf Final Demand Goods} (4935.4)} \\ \hline
\multicolumn{3}{|c|}{{\sf Investment Goods} (2352.5)}\\
\hline
Capital Goods (1662.1) &1& Manufacturing Equipment (530.7)\\
&2& Electricity (148.1)\\
&3&Communication and Broadcasting (48.8)\\
&4&Agriculture (31.0)\\
&5& Construction (129.6)\\
&6&Transport (381.3)\\
&7&Offices (175.4)\\
&8&Other Capital Goods (217.2)\\
\hline
Construction Goods (690.4)&9&Construction (568.1)\\
&10&Engineering (122.3)\\
\hline
\multicolumn{3}{|c|}{{\sf Consumer Goods} (2582.9)}\\
\hline
Durable Consumer&11& House Work (62.3)\\
Goods (1267.9)&12& Heating/Cooling Equipment (62.5)\\
&13&Furniture \& Furnishings (43.4)\\
&14&Education \& Amusement (246.5)\\
&15&Motor Vehicles (853.2)\\
\hline
Nondurable &16& House Work (649.7)\\
Consumer Goods (1315.0)&17& Education \& Amusement (105.2)\\
&18& Clothing \& Footwear (92.2)\\
&19& Food \& Beverage (467.9)\\
\hline\hline
\multicolumn{3}{|>{\columncolor[gray]{0.6}}c|}{{\sf Producer Goods} (5064.6)}\\ 
\hline
&20& Mining \& Manufacturing (4601.7)\\
&21& Others (462.9)\\
\hline
\end{tabular}
\end{table*}

Figure~\ref{fig:IIPaverage} shows temporal change of the averaged IIP data for production, shipments, and inventory during the period of January 1988 to June 2009. The ongoing global recession is traced back to the subprime mortgage crisis in the U.S., which became apparent in 2007. The economic shock has affected Japan without exception, leading to a dramatic drop in the production activities of the country, as shown in the figure. 

\figp{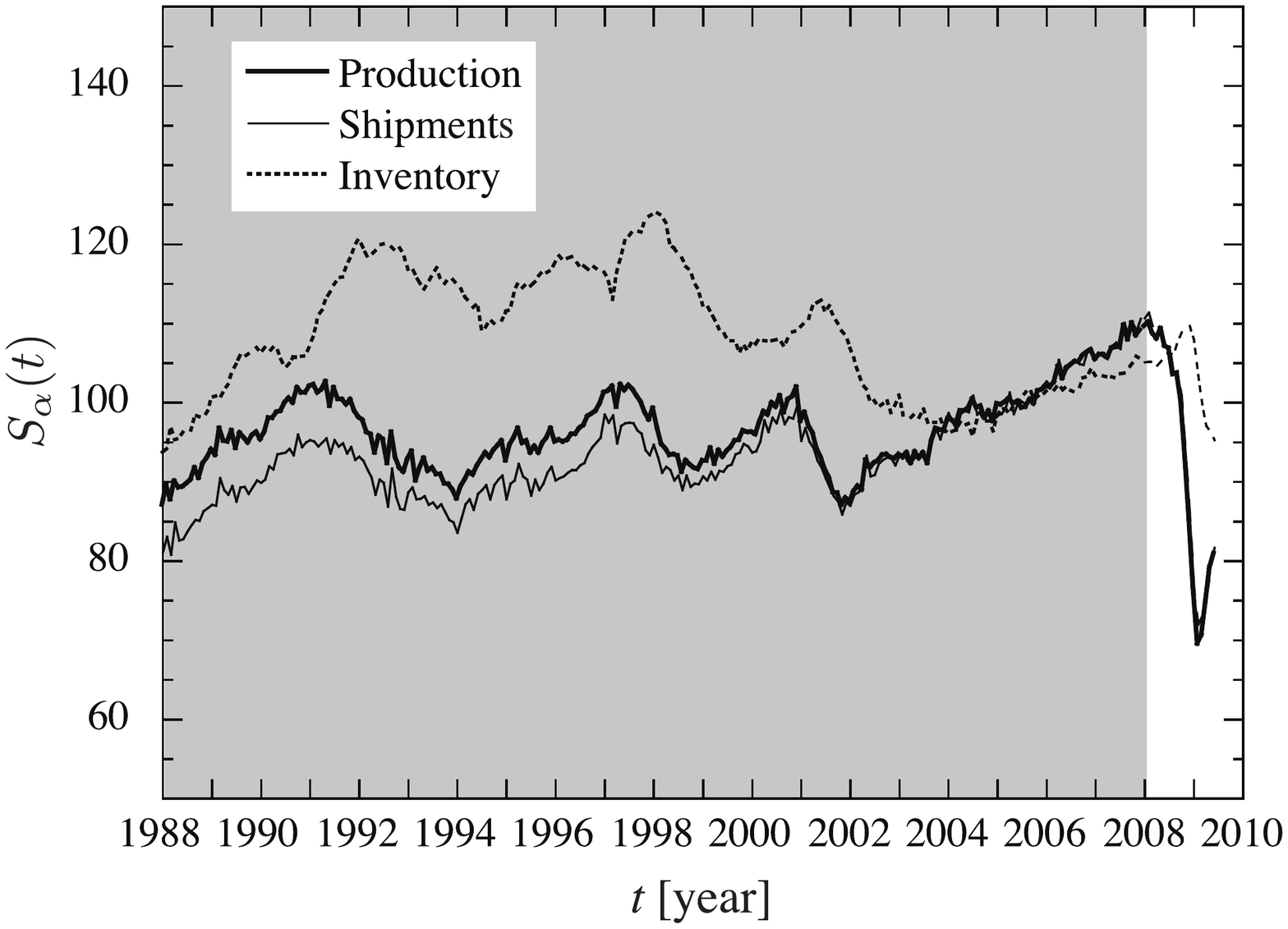}{Averaged IIP data $S_\alpha$ for production (thick solid line), shipment (thin solid line), and inventory (dotted line) as a function of time $t$. The correlation matrix is calculated using the data in the gray shaded area from January 1988 to December 2007.}{0.4}
{8.75732 216.296 558.309 611.39}

\def\ga{_{\alpha,g}}
\def\tn{N'}
Since some of the entries, such as $g=16$ and $17$, are missing before January 1988, we use the data~\citep{iip}
for the 240 months from January 1988 to December 2007. Further, this chosen period for the study excludes the abnormal behavior of the IIP data due to The Great Recession.
We denote the IIP data for goods as $S\ga(t_j)$,
where $\alpha=1,2,$ and $3$ for production (value added), shipments, 
and inventory, respectively.
Similarly, $g=1,2, \dots, 21$ denotes the 21 categories of goods, and 
$t_j=j \Delta t$ with $\Delta t=1$ month and $j=1,2, \dots, N (=240)$; $j=1$ and $j=N$ correspond to 1/1988 and 12/2007, respectively.
The logarithmic growth rate $r\ga(t_j)$ 
is defined as
\begin{equation}
r\ga(t_j):=\log_{10}\!\left[ \frac{S\ga(t_{j+1})}{S\ga(t_j)}\right],
\label{logreturn}
\end{equation}
where $j$ runs from 1 to $\tn:=N-1(=239)$. Then, it is normalized as
\begin{equation}
w\ga(t_j):= \frac{r\ga(t_j)-\langle r\ga\rangle_t}{\sigma\ga}\,
\label{wnormal}
\end{equation}
where $\langle \cdot \rangle_t$ denotes average over time $t_1, \dots, t_{\tn}$
and $\sigma\ga$ is the standard deviation of $r\ga$ over time.
Definition (\ref{wnormal}) ensures that the set $w_{\ga} := \{w\ga(t_1), w\ga(t_2), \dots , w\ga(t_{\tn})\}$
has an average of zero and a standard deviation of one.

Figure~\ref{fig:w2} shows an overview of how the volatility $w^{2}\ga(t_j)$ of the standardized IIP data behaves on a time-goods plane. Unfortunately, the visualization does not allow for detecting any correlations involved in the IIP data. One may even doubt whether useful information on interindustrial relations truly exists in the data.

\figp{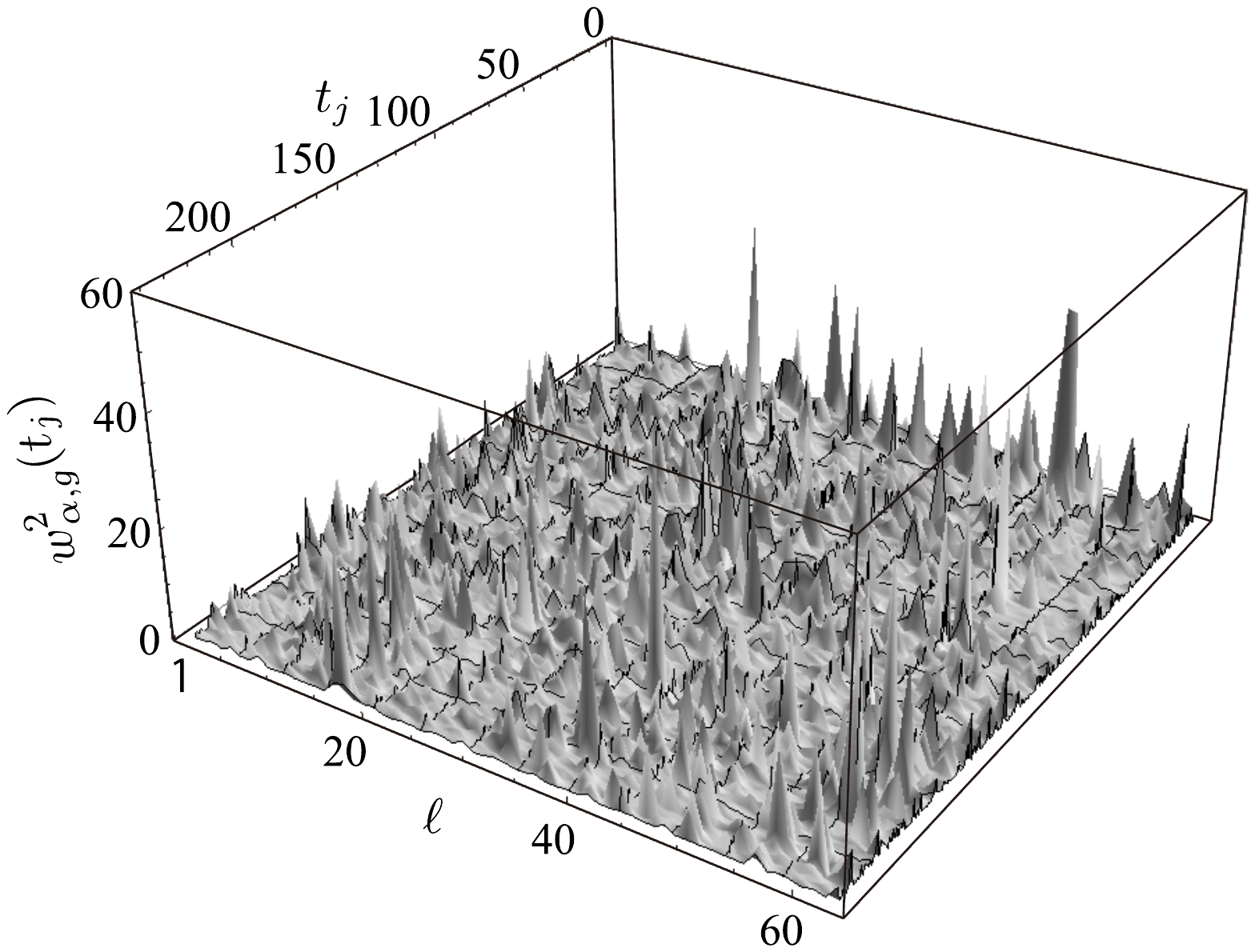}{Bird's-eye view of volatility of standardized IIP data in a panel form,  where the index $\ell$ is defined as $\ell:=21(\alpha-1)+g$.}{0.4}
{56.3618 249.837 503.25 587.703}

\def\cc{\bm{C}}
To answer the obvious question that would arise here, we begin with calculating the equal-time correlation matrix $\cc$ of $\{ w_{\ga} \}$ according to\begin{equation}
C_{\alpha,g;\beta,h}=\langle w_{\alpha,g}(t) w_{\beta,h}(t) \rangle_t,
\label{Cdef}
\end{equation}
whose diagonal elements are unity by definition of the normalized growth rate $w\ga (t_{j})$.
Since $\alpha$ ($\beta$) runs from 1 to 3 and $g$ ($h$) runs
from 1 to 21, the matrix $\cc$ has $M\times M$ ($M=63$) components.
We denote the eigenvalues and the corresponding eigenvectors of the correlation matrix as $\lambda^{(n)}$
and $\bm{V}^{(n)}$, respectively:
\begin{equation}
\bm{C}\,\bm{V}^{(n)} = \lambda^{(n)} \bm{V}^{(n)},
\end{equation}
where the eigenvalues are sorted in descending order of their values and the norm of eigenvectors is set to unity.

On the basis of the eigenvectors $\bm{V}^{(n)}$ thus obtained, the normalized growth rate $w\ga(t_j)$ can be decomposed into
\begin{equation}
w\ga(t_j)=\sum_{n=1}^{M} a_n(t_j)\, V\ga^{(n)}.
\label{d1}
\end{equation}
The correlation matrix $\cc$ is also decomposable in terms of the eigenvalues and eigenvectors as
\begin{equation}
\bm{C}=\sum_{n=1}^{M}\lambda^{(n)}\bm{V}^{(n)}\bm{V}^{(n){\rm T}}.
\label{CVV}
\end{equation}
The eigenvalues satisfy the following trace constraint:
\begin{equation}
\sum_{n=1}^{M}\lambda^{(n)}=M.
\label{tracec}
\end{equation}
By substituting Eq.~(\ref{d1}) into Eq.~(\ref{Cdef}) and comparing it with Eq.~(\ref{CVV}),
we find that
\begin{equation}
\langle a_n(t) a_{n'}(t) \rangle_t = \delta_{nn'}\lambda^{(n)}.
\label{lamaa}
\end{equation}
The eigenvalue of each eigenmode thus represents the strength of fluctuations associated with the mode. Figure~\ref{fig:cj2} shows the temporal variation of $a^{2}_{n}(t)$, which is in sharp contrast to the results shown in Fig.~\ref{fig:w2}. The transformation of the base for describing the IIP data reveals that very few degrees of freedom actually are responsible for the complicated behavior of the IIP. 

\figp{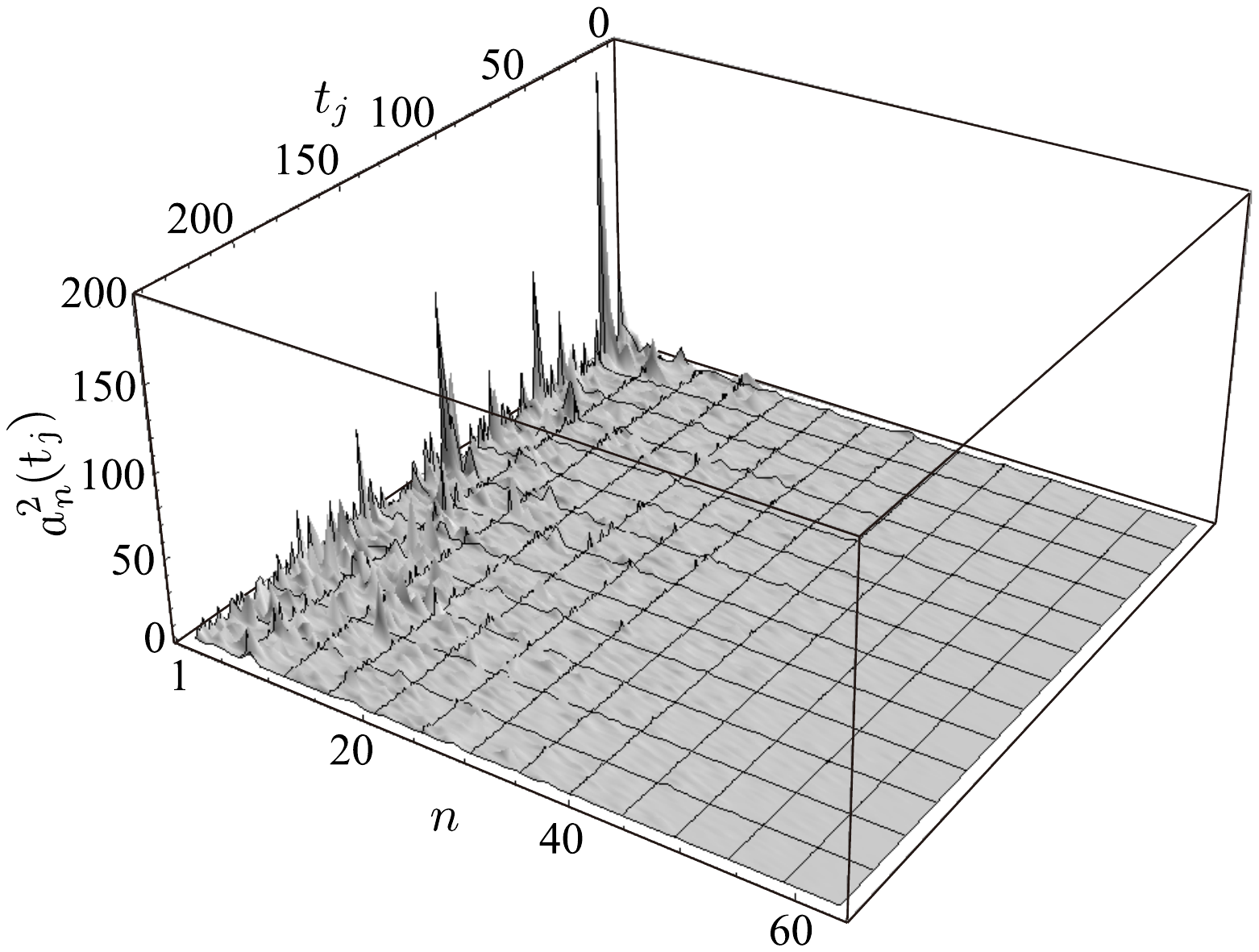}{Temporal variation of strength of fluctuations associated with each eigenmode, where $n$ is an index assigned to eigenmodes in descending order of their eigenvalues.}{0.4}
{58.1836 251.837 503.545 587.152}

\figp{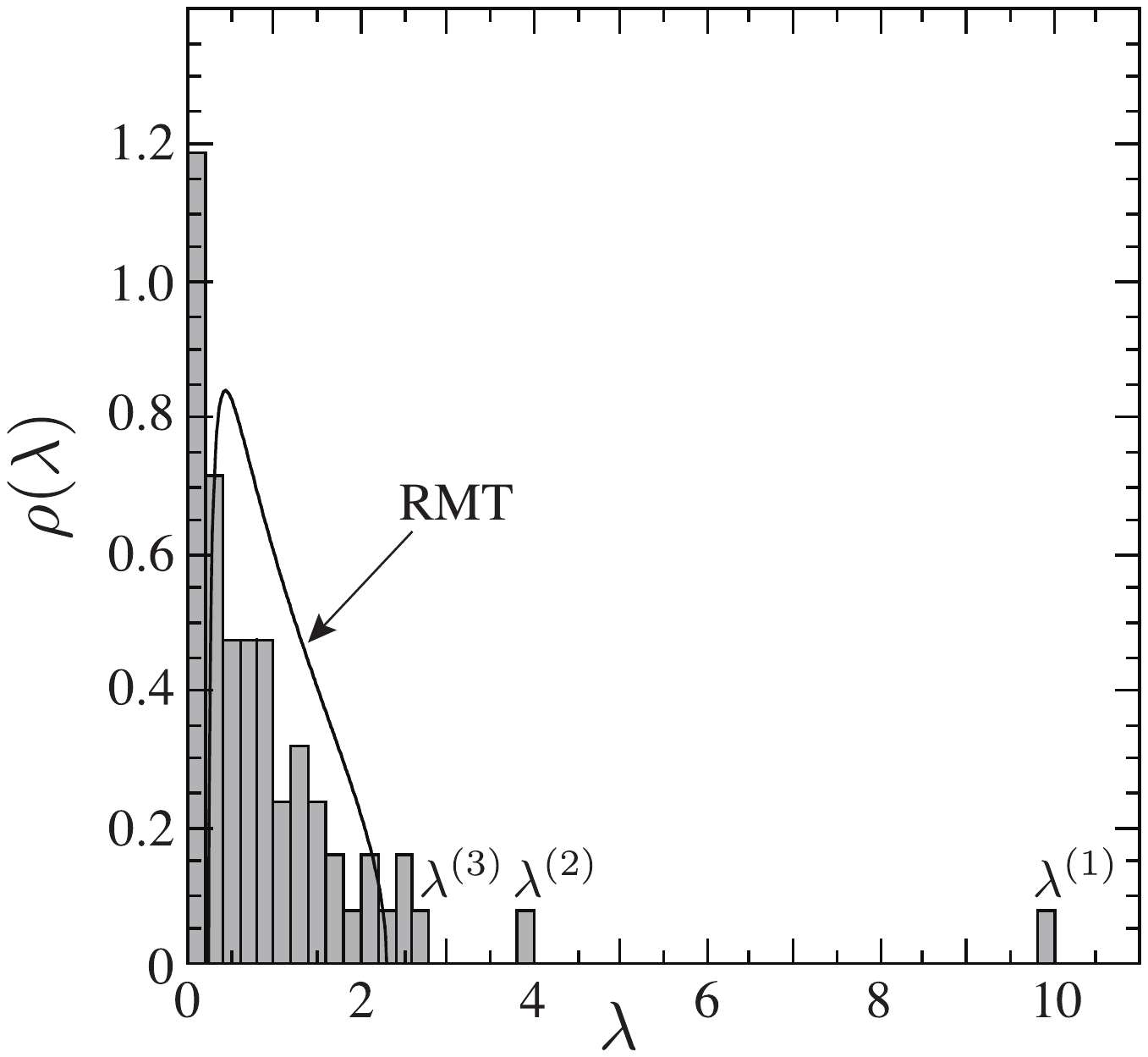}{Probability distribution function $\rho(\lambda)$ for eigenvalues ($\lambda$) of correlation matrix derived from IIP data, in comparison with corresponding result of random matrix theory (RMT) represented by solid curve.}{0.35}
{243.466 64.7803 629.5 419.775}

Thanks to the RMT, we are able to quantify how many eigenmodes should be considered. Probability distribution function $\rho(\lambda)$ for the eigenvalues ($\lambda$) of the correlation matrix $\bm{C}$ is shown in Fig.~\ref{fig:dist_egv}. It is compared with the corresponding result~\citep{sengupta1999dsv} of the RMT in the limit of infinite dimensions:
\begin{equation}
\rho(\lambda)=
\begin{cases}\displaystyle
\frac{Q}{2\pi}
\frac{\sqrt{(\lambda_+-\lambda)(\lambda-\lambda_-)}}{\lambda}
& \mbox{for } \lambda_- \le \lambda \le \lambda_+, \\[10pt]
0 & \mbox{otherwise,}
\end{cases}
\label{rmte1}
\end{equation}
where $Q:=\tn/M \simeq 3.79\ (>1)$ and the upper and lower bounds $\lambda_\pm$ for $\lambda$ are given as
\begin{equation}
\lambda_\pm=\frac{(1\pm\sqrt{Q})^2}{Q} \simeq
\begin{cases}
2.29\\
0.237
\end{cases}
.
\label{rmte2}
\end{equation}
We see that the largest and the second largest eigenvalues, designated as $\lambda^{(1)} (\simeq9.95)$ and $\lambda^{(2)} (\simeq3.83)$, are well separated from the eigenvalue distribution predicted by the RMT, whereas the third largest eigenvalue $\lambda^{(3)} (\simeq2.77)$ is adjacent to the continuum. Therefore, only 2 eigenmodes out of a total of 63 are of statistical significance according to the RMT. 

Readers may be curious about the present construction of a correlation matrix by mixing up data for production, shipments, and inventory, because these are very different species of data at first glance. Thus far, physicists have applied the RMT mainly to analyses of stock data having similar characteristics.
In this sense our approach is quite radical. However, production, shipments, and inventory form a trinity in the economic theory for business cycles, so that those variables should be treated on an equal footing. Using the two dominant eigenmodes, in fact, we were successful in proving the existence of intrinsic business cycles~\citep{bc}.

In passing, we note that one may favor the growth rate itself defined by
\begin{equation}
r\ga(t_j):=\frac{S\ga(t_{j+1})-S\ga(t_j)}{S\ga(t_j)}
\label{growth_rate}
\end{equation}
for the present analysis over the logarithmic growth rate~(\ref{logreturn}). If the relative change in $S\ga(t_j)$ is small, we need not distinguish between Eqs. (\ref{logreturn}) and (\ref{growth_rate}) numerically. To confirm that the results obtained here are insensitive to the choice of stochastic variables, we repeated the same calculation by using Eq. (\ref{growth_rate}) and found no appreciable difference between the two calculations for the dominant eigenvalues and their associated eigenvectors. For instance, the first three largest eigenvalues 9.95, 3.83, and 2.77 as shown in Fig.~\ref{fig:dist_egv} are replaced with 9.96, 3.73, and 2.78, respectively.

\section{Rotational Random Shuffling}\label{sec:rs}
There are two major sources of noise in the IIP data. One of them, corresponding to thermal noise in physical systems, arises from elimination of a large number of degrees of freedom from our scope as hidden variables. This highlights the stochastic nature of the IIP and has a strong influence on autocorrelation of all goods. The other source of noise originates from the finite length of time series data. Such statistical noise hinders the detection of correlations among different goods in the IIP data. If one could have data of infinite length, statistical noise would disappear in the mutual correlations and only thermal noise would remain. These two types of noise should be distinguished conceptually. The RMT is an effective tool for eliminating statistical noise from raw data to extract genuine mutual correlations. 


However, the noise reduction method based on the RMT heavily depends on the following assumption: stochastic variables would be totally independent if correlations between different variables were switched off. Such a null hypothesis simultaneously excludes both autocorrelations and mutual correlations. In the case of daily change in Japanese stock prices that were available~\citep{nakayamai2009rmt} to us, we found no detectable autocorrelations in the corresponding variables; therefore, the RMT functions ideally. In contrast, the IIP data have significant autocorrelations as shown in Fig.~\ref{fig:autocorrelation}, where the autocorrelation function $R_{\alpha, g}(t)$ of the normalized 
growth rate $w_{\alpha,g}$ is defined as
\begin{equation}
R_{\alpha, g}(t_m):=\frac{1}{N'-m}\sum_{j=1}^{N'-m} w_{\alpha,g}(t_j) w_{\alpha,g}(t_{j+m}).
\label{auto1}
\end{equation}
By definition, $R_{\alpha, g}(0)=1$, and 
if there are no autocorrelations, $R_{\alpha, g}(t_m)=0$ for $m \ge 1$.
We observe that both production ($\alpha=1$) and shipments ($\alpha=2$) have
nontrivial values of autocorrelations at $t=1$ month, whereas there is no clear evidence for autocorrelations for inventory ($\alpha=3$) in the same time interval; the values averaged over 21 goods are $R_1(1) \simeq -0.31$, $R_2(1) \simeq -0.39$, and $R_3(1) \simeq 0.007$. Beyond the one-month time lag, however, we find no appreciable autocorrelations for any of these three categories.

\figp{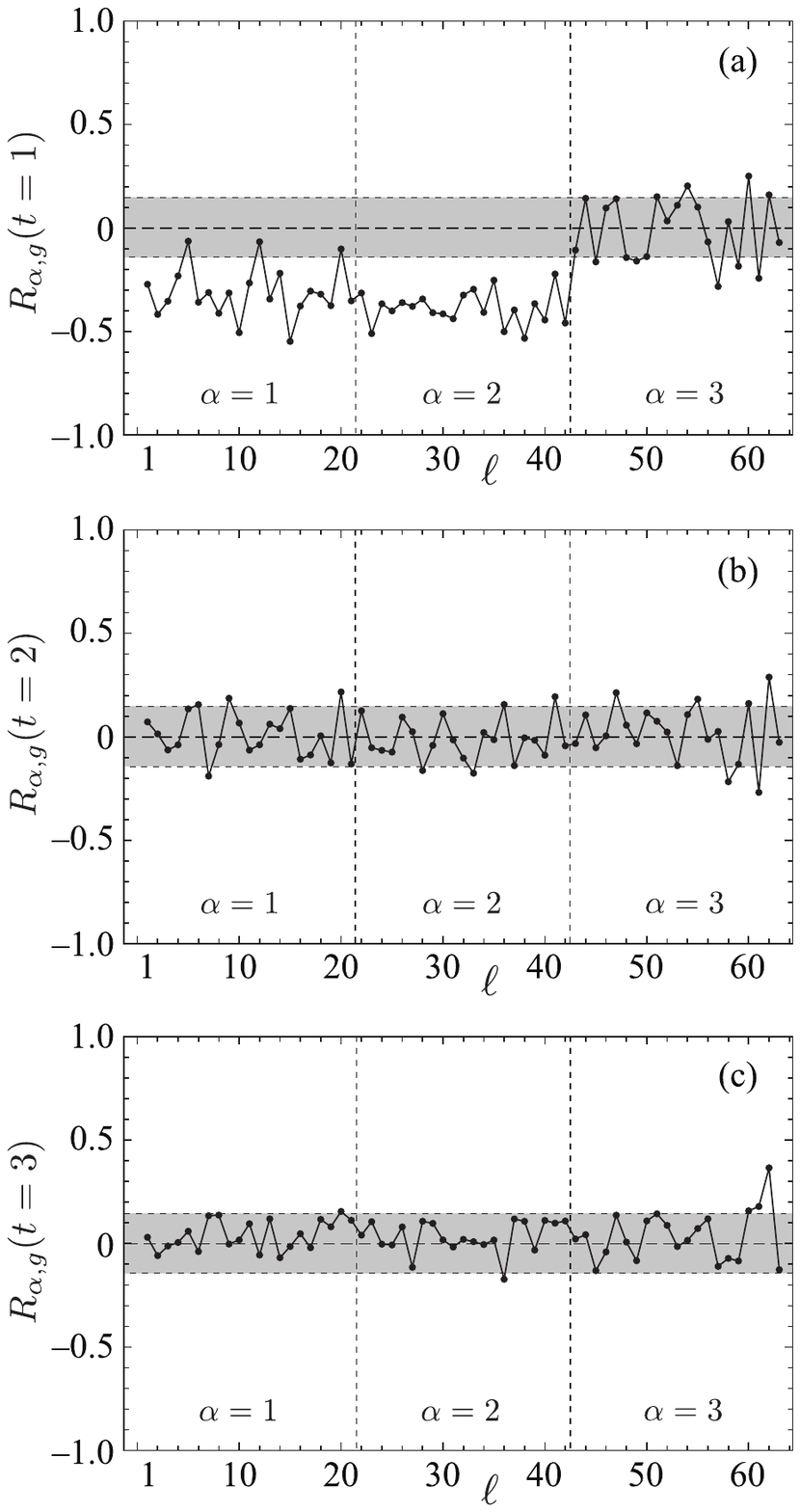}{Autocorrelation functions $R_{\alpha,g}(t)$ of production ($\alpha=1$), shipments ($\alpha=2$), and inventory ($\alpha=3$) for each of the goods ($g=1,2,...,21$) at $t=1$~(a), $t=2$~(b), and $t=3$~(c), where the index $\ell$ on the horizontal axis is defined in the same way as in Fig.~\ref{fig:w2}. The 95\% confidence level for no autocorrelations is represented by the gray shaded band in each panel.}{0.35}
{166.636 168.093 429.785 670.739}

\figp{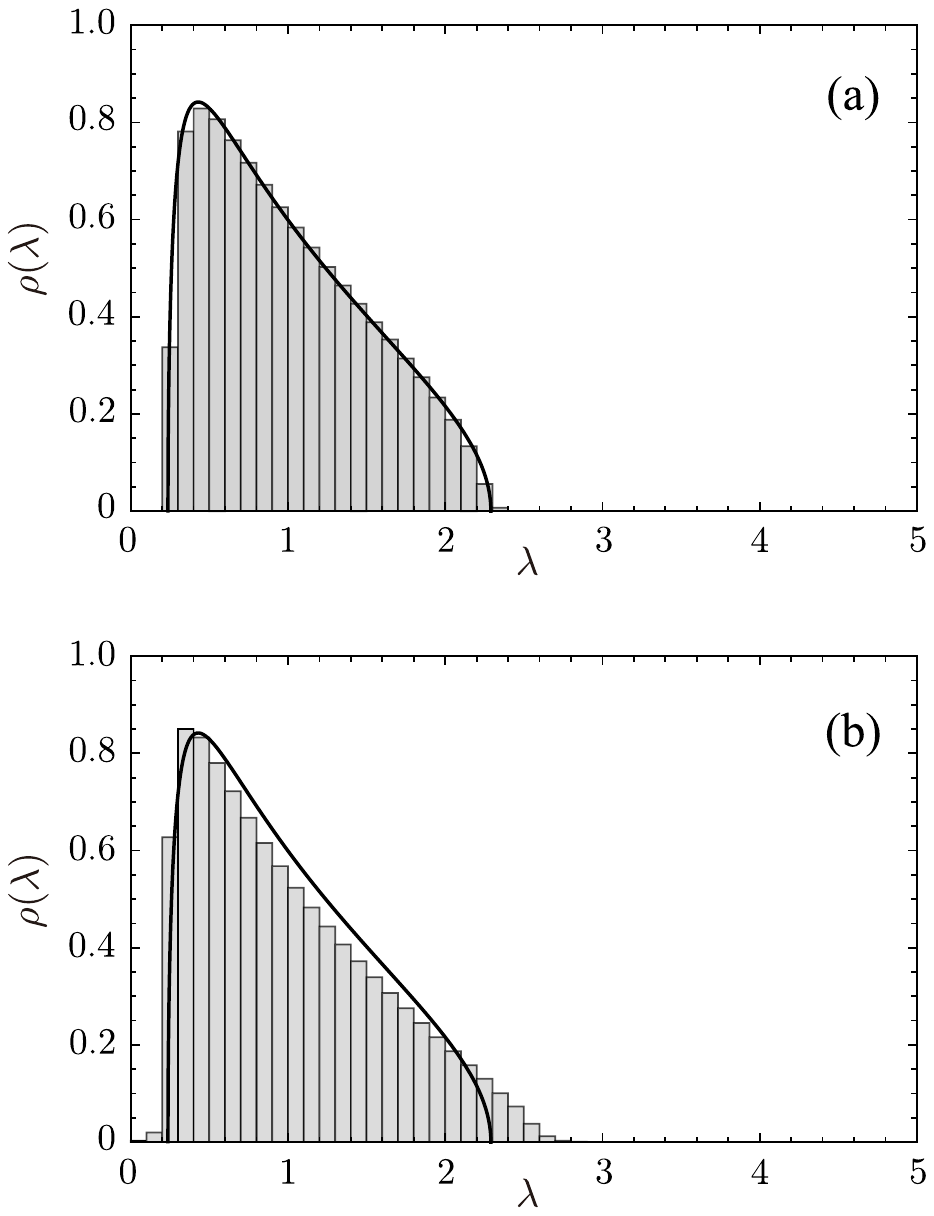}{Same as Fig.~\ref{fig:dist_egv}, but for eigenvalues ($\lambda$) of the correlation matrix obtained by shuffling IIP data completely (a) or rotationally (b) in the time direction. It should be noted that the autocorrelations involved in the IIP data are preserved in rotational shuffling.}{0.35}
{158.662 284.849 423.588 630.03}

To formulate the null hypothesis of the RMT using actual data, one may shuffle the IIP data completely in the time direction. In fact, the eigenvalue distribution of the resulting correlation matrix reduces to that of the RMT as demonstrated in panel (a) of Fig.~\ref{fig:dist_egv_shuffle}, where a total of $10^{5}$ samples were generated. Departure from the RMT owing to finiteness of the data size is almost negligible even for such small-scale data as the IIP. This randomization process inevitably destroys both autocorrelations and mutual correlations. From a methodological point of view, it is favorable to deal with these two types of correlations separately. 

We instead propose to shuffle the data rotationally in the time direction, imposing the following periodic boundary condition on each of the time series:
\begin{equation}
w_{\alpha,g}(t_j)\rightarrow w_{\alpha,g}(t_{{\rm Mod}(j-\tau,N')}),
\end{equation}
where $\tau\in[0,N'-1]$ is a (pseudo-)random integer and is different for each $\alpha$ and $g$. This randomization destroys only the mutual correlations involved in the data, with the autocorrelations left as they are; therefore, it provides us with a null hypothesis more appropriate than that of the RMT. 

Panel (b) of Fig.~\ref{fig:dist_egv_shuffle} shows the result in rotational shuffling with the same number of samples as that in the complete shuffling. We find that the existence of autocorrelations alone leads to departure from the RMT. The third largest eigenvalue $\lambda^{(3)} \simeq 2.77$ becomes even closer to the upper limit, $\lambda'_{+}=2.47 \pm 0.20$, of the eigenvalues obtained on the basis of the alternative null hypothesis, where the error is estimated at 95\% confidence level. This result reinforces neglect of the third eigenmode by the RMT.

Thus, this new method for data shuffling conceptually clarifies noise elimination for the correlation matrix, although the difference in the eigenvalue distribution from that of the RMT is practically not very dramatic. In addition, we note that the rotational shuffling of the stock price data in Japan reproduces the RMT result quite well, as is expected from the fact that no appreciable autocorrelations are observed there.

\section{Genuine Correlation Matrix}\label{sec:gcm}
\def\ctrue{\bm{C}^{\rm (G)}}
In the current system of IIP data, the above careful arguments permit us to adopt 
\begin{equation}
\ctrue:=\sum_{n=1}^{2}\lambda^{(n)}\bm{V}^{(n)}\bm{V}^{(n){\rm T}}+\mbox{[diagonal terms]}
\label{eq:C_truemtx}
\end{equation}
as a genuine correlation matrix, which consists of just the first and second eigenvector components in the spectral representation (\ref{CVV}) of $\bm{C}$ plus the diagonal terms, thereby ensuring that all the diagonal components are 1. We note that self-correlations of stochastic variables always exist even if they are merely noise.
The components of $\bm{C}$ are explicitly written as  
\begin{equation}
{C}^{\rm (G)}_{\ell m}=
\begin{cases}
1 & \mbox{for } \ell=m,\\
\displaystyle
\sum_{n=1}^{2}\lambda^{(n)}{V}^{(n)}_\ell {V}^{(n)}_m & \mbox{otherwise}.
\end{cases}
\label{eq:C_true}
\end{equation}

The eigenvectors $\bm{V}^{(1)}$ and $\bm{V}^{(2)}$, associated with $\lambda^{(1)}$ and $\lambda^{(2)}$, are shown in panels (a) and (b) of Fig.~\ref{fig:egvec12}, respectively. These two eigenvectors have characteristic features that distinguish them from each other. The eigenvector $\bm{V}^{(1)}$ represents an economic mode in which production and shipments of all goods expand (shrink) synchronously with decreasing (increasing) inventory of producer goods. This corresponds to the market mode obtained for the largest eigenvalue in the stock market analyses~\citep{laloux1999ndf, plerou1999uan}, and may be referred to as the ``aggregate demand" mode according to Keynes' principle of effective demand: both shipments and production in all the sectors are moved jointly by aggregate demand~\citep{Keynes1936}. On the other hand, the eigenvector $\bm{V}^{(2)}$ is a mode that apparently represents dynamics of inventory, i.e., accumulation or clearance of inventory, for most goods, including producer goods. We further find positive correlation between production enhancement and inventory accumulation for most goods. This finding indicates that production has a kind of inertia in its response to change of demands.

\figp{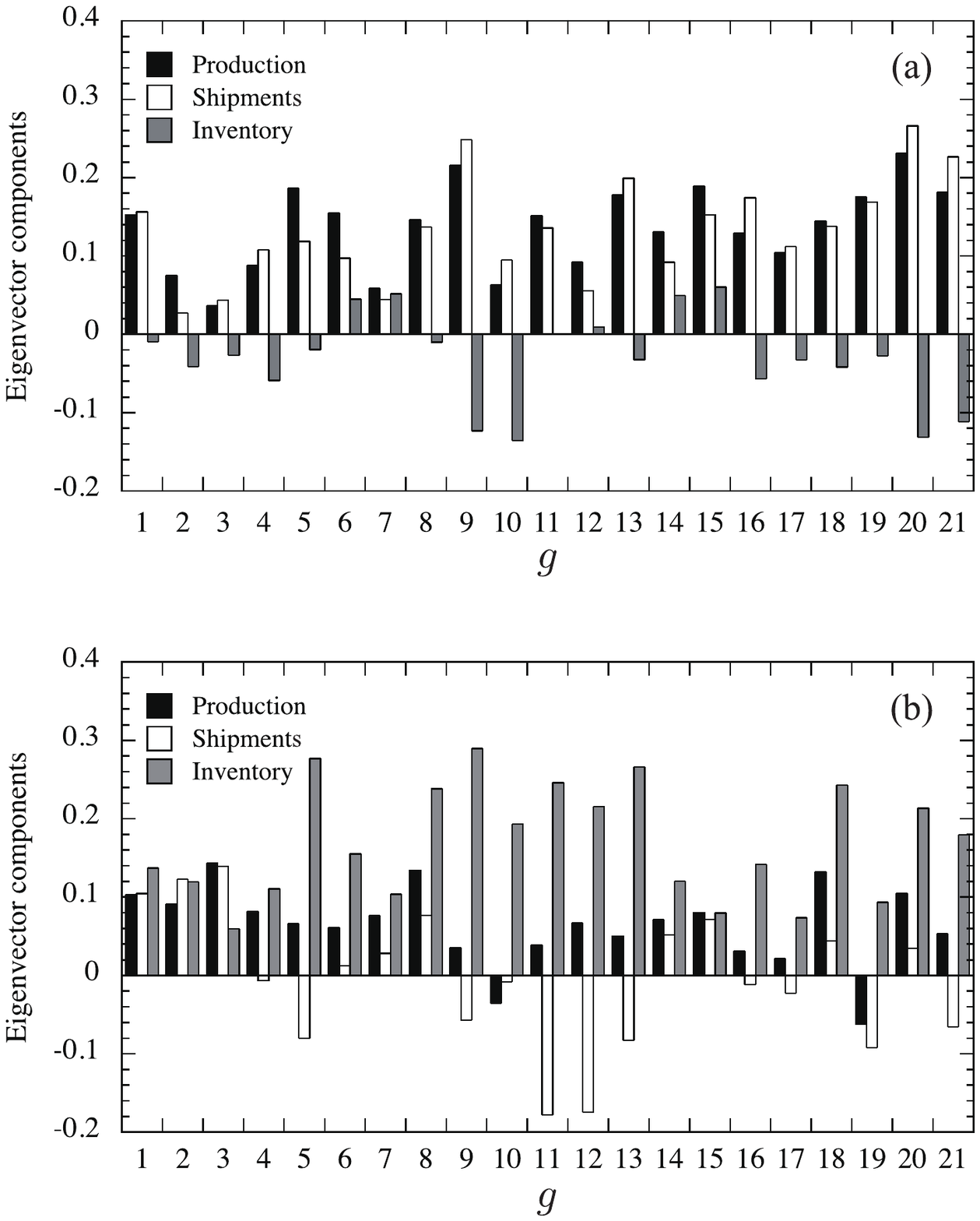}{Eigenvector components corresponding to largest (a) and second largest (b) eigenvalues for correlation matrix of IIP time-series data.}{0.4}
{94.9165 201.644 510.501 723.842}

Accordingly, we project out raw fluctuations of $w_\ell(t_{j})$ onto the first and second eigenmodes; that is, only the first two terms are retained and the remaining terms are regarded as just noise in expansion (\ref{d1}):
\begin{equation}
w_{\ell}(t_j)=\sum_{n=1}^{2} a_n(t_j)\, V_{\ell}^{(n)} + \textrm{[noise]}.
\label{d1m2}
\end{equation}
This process extracts statistically meaningful information on mutual correlations among $w_\ell(t_{j})$ as has been already discussed in Secs.~\ref{sec:rmt-iip} and \ref{sec:rs}. Collaboration of these two modes results in inherent business cycles with periods of 40 and 60 months throughout the economy. The cycles are accounted for by time lags in information flow between demand of goods by consumers and decision making of firms on production~\citep{bc}; inventory fills this information gap. If we singled out the most dominant mode alone in Eq.~(\ref{d1}), all $w_\ell(t_{j})$ would oscillate without phase difference. As will be shown later, each goods possesses its own characteristics in the phase relations among production, shipments, and inventory.

Temporal change of the two principal factors $a_{1}(t)$ and $a_{2}(t)$ is plotted in panel (a) of Fig.~\ref{fig:cjmv1}. Since the functional behavior of these variables is very noisy, we take their simple moving average defined as
\begin{equation}
\overline{a_n(t_j)}:=\frac{1}{2\xi +1}\sum_{k=-\xi}^{\xi} a_n(t_{j+k}),
\label{mvav_a}
\end{equation}
where $\xi$ is a characteristic time scale for smoothing. This process eliminates ``thermal" noise present in the original data. Actually, the moving average was taken with $\xi=6$; the results for $\overline{a_{1}(t)}$ and $\overline{a_{2}(t)}$ are shown in panel (b) of Fig.~\ref{fig:cjmv1}. We see that the moving-average operation significantly reduces the level of noise present in $a_{n}(t)$. It is noteworthy that Fig.~\ref{fig:cjmv1} indicates the existence of some mechanical relationship between $a_{1}(t)$ and $a_{2}(t)$. This finding is ascertained more quantitatively from Fig.~\ref{fig:correlation}, in which the correlation coefficient between $\overline{a_1(t)}$ and $\overline{a_2(t-\tau)}$,
\begin{equation}
C_{\overline{a_{1}}\;\overline{a_{2}}}(\tau)=\frac{\Bigl\langle \overline{a_{1}(t)} \; \overline{a_{2}(t-\tau)} \Bigr\rangle_{t}}{ \sqrt{\Bigl\langle \overline{a_{1}(t)}^2 \Bigr\rangle_{t} \, \Bigl\langle \overline{a_{2}(t)}^2 \Bigr\rangle_{t}}}\, ,
\end{equation}
is plotted as a function of time lag $\tau$. A correlation as large as 0.7 is detected between the two dominant modes around $\tau=10$ months. Detailed study of the underlying dynamics in the economic system is in progress and will be reported elsewhere.

\figp{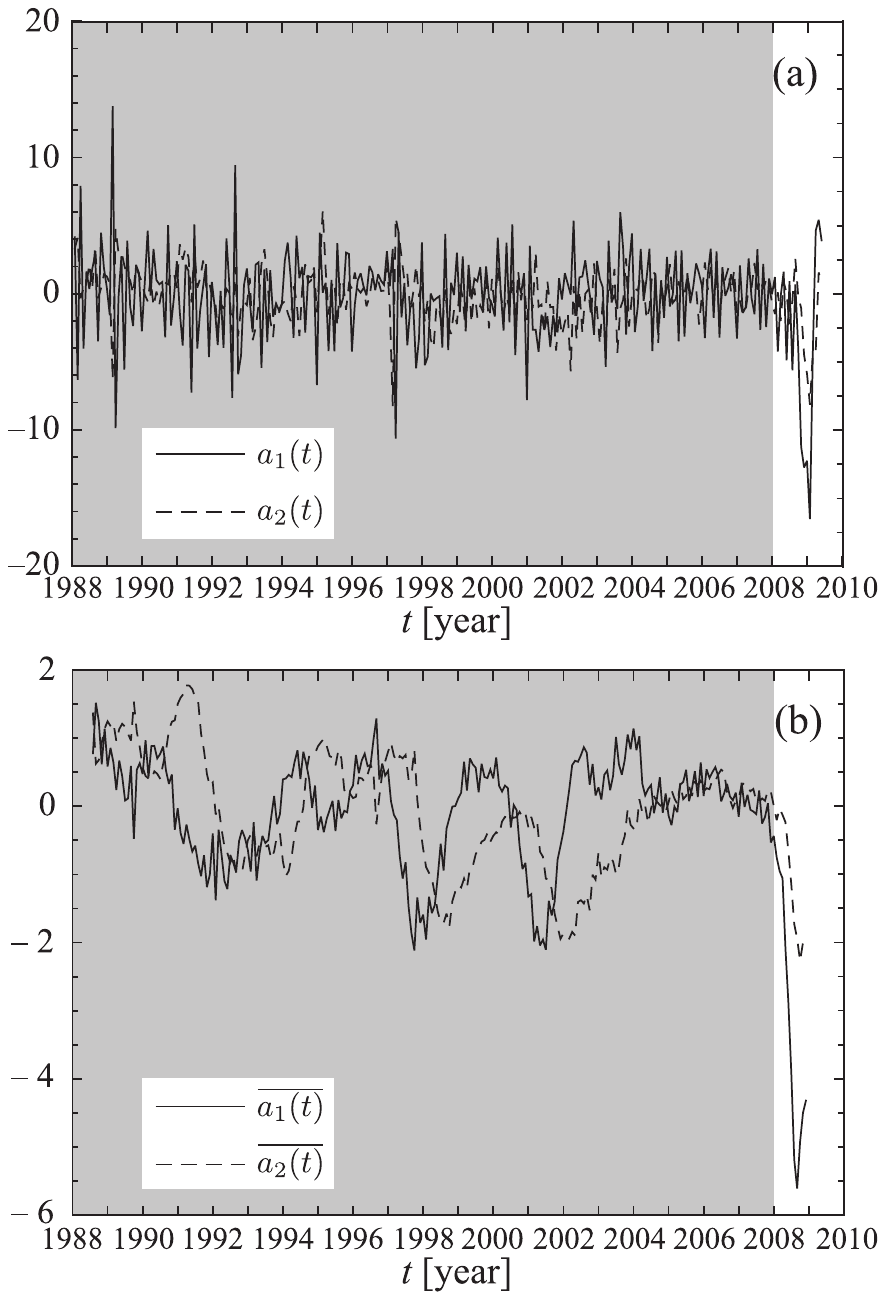}{Two principal factors $a_{1}$ and $a_{2}$ as a function of time $t$. Panel (a) shows the originally obtained results, and panel (b) shows those smoothed by the moving-average operation (\ref{mvav_a}) with $\xi=6$. The eigenmodes for business fluctuations were determined using the data in the gray shaded area.}{0.4}
{168.912 321.953 419.396 692.147}

\figp{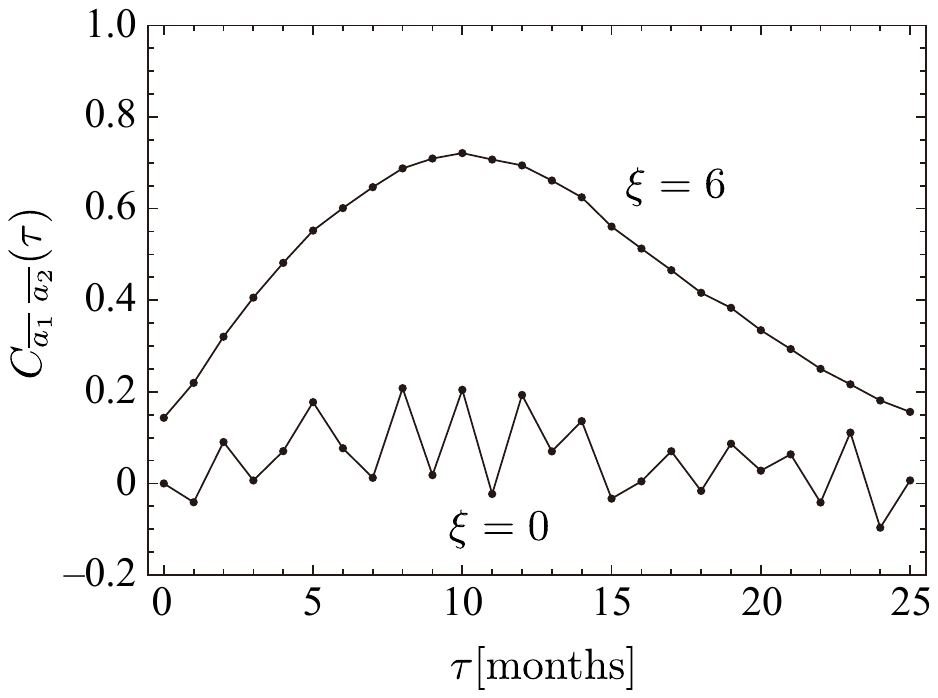}{Correlation coefficient $C_{\overline{a_{1}}\;\overline{a_{2}}}(\tau)$ between $\overline{a_1(t)}$ and $\overline{a_2(t-\tau)}$ as a function of time lag $\tau$, calculated during the normal period (from January 1988 to December 2007) with $\xi=6$ for the moving-average operation. A comparison of this result with that obtained using unsmoothed data with $\xi=0$ is also shown in the figure.}{0.35}
{152.729 311.175 418.075 507.148}

\section{Fluctuation-Dissipation Theory}\label{sec:fdt}
The fluctuation-dissipation theory plays a central role in nonequilibrium statistical mechanics because this theory establishes a general relation between fluctuation properties of a physical system in equilibrium and response properties of the system to small external perturbations. We assume that the theory still is applicable to the economic system under study here. This assumption provides us with a framework to derive input-output interindustrial relations in the system. Its validity in view of how the system responded to the recent economic crisis will be discussed later .

Let us denote our variable $w_{\alpha, g}$ as $w_\ell$, whose
index $\ell :=21(\alpha-1)+g$ runs from 1 to 63.
We assume that $w_\ell$ obeys dynamics governed by a Hamiltonian
$H(\{w\}, \{x\})$, where $\{x\}$ is a set of ``hidden" variables in the 
system, which encompass all variables in the
current economics. These numerous variables interact with
each other in a nonlinear chaotic way and hence,
the temporal change of $w_\ell$ appears stochastic in the same way as that of a Brownian particle. 
The existence of underlying dynamics in the IIP~\citep{bc}, as demonstrated in Figs.~\ref{fig:cjmv1} and \ref{fig:correlation}, strongly supports this idea borrowed from mechanics of motion.

Actually, however, the economy of a nation is quite open now; therefore, it could potentially be subjected to perturbations such as disasters, political issues, and trade issues. We thus add external forces $\epsilon_\ell (t)$ to the system; then, the total Hamiltonian ${\cal H}$ becomes
\begin{equation}
{\cal H}=H(\{w\}, \{x\})-\sum_{\ell=1}^M \epsilon_\ell (t) w_\ell .
\label{eq:total_H}
\end{equation}
This extra term represents external perturbations to the equation of motion for $w_\ell$:
\begin{align}
\frac{d p_\ell}{dt}&=-\frac{\partial H}{\partial w_\ell }+\epsilon_\ell (t),\\
\frac{d w_\ell}{dt}&=\frac{\partial H}{\partial p_\ell },
\end{align}
where $p_\ell$ is the momentum conjugate to $w_\ell$. Therefore, $\epsilon_\ell(t)$ directly affects $w_\ell$ at time $t$,
the effect of which then extends to other $w$'s through direct and indirect interactions among them. 

For simplicity, let us assume that $\{\epsilon\}$ is constant in time. Thus, the perturbation set $\{\epsilon\}$ induces a static shift of the equilibrium positions of 
the variables $\{ w \}$, which otherwise move stochastically
around the origin $w_\ell=0$. If the perturbation is weak, the shift $\langle w_\ell\rangle$ thus induced can be expressed by the following linear response relation:
\begin{equation}
\langle w_\ell\rangle =\sum_{m=1}^M \chi_{\ell m}\epsilon_m,
\label{eq:linear_response}
\end{equation}
where the ensemble average denoted by $\langle \cdot \rangle$ has replaced the time average.
The coefficients $\{\chi\}$ are the result of the interactions, and they are 
called ``magnetic susceptibility" while describing the physics of magnetic materials.

Once such a set of susceptibilities is available, we can quantify the response of the economic system to external perturbations. For instance, suppose that the government adopts an economic policy to increase the shipment of one of the final demand goods by $\langle w_m \rangle$ with a stimulus $\epsilon_m$. The resulting changes in production, shipments, and inventory of goods are given as
\begin{equation}
\begin{split}
\langle w_1\rangle &=\chi_{1m}\epsilon_m, \\
\vdots \\
\langle w_M\rangle &=\chi_{Mm}\epsilon_m. \\
\end{split}
\label{wxe}
\end{equation}
Since $\epsilon_m$ is not an observable quantity, it should be appropriate to eliminate $\epsilon_m$ appearing in Eq.~(\ref{wxe}) and express ripple effects on the economy in terms of $\langle w_m\rangle$ as
\begin{equation}
\begin{split}
\langle w_1\rangle &=\frac{\chi_{1m}}{\chi_{mm}}\langle w_m \rangle, \\
\vdots \\
\langle w_M\rangle &=\frac{\chi_{Mm}}{\chi_{mm}}\langle w_m\rangle. \\
\end{split}
\label{wxw}
\end{equation}

In Sec.~\ref{sec:ic}, we demonstrate that Eq. (\ref{wxw}) can provide quantitative information on input-output interindustrial relations in Japan. Further, one may make reverse use of the linear response relation (\ref{eq:linear_response}) to distinguish and detect external perturbation from observed economic changes in $\{ w \}$; this is discussed in Sec.~\ref{sec:iir}.

Now, the remaining problem is how to calculate $\{\chi\}$. To this end, we invoke the concept of the fluctuation-dissipation (FD) theorem in statistical physics. If we assume that the stochastic process of $\{ w \}$ is characterized by Gibbs' ensemble, then the probability density function (PDF) $P(\{ w \},\{ \epsilon \})$ for $\{ w \}$ is given as 
\begin{equation}
P(\{ w \},\{ \epsilon \}) \propto \exp [-\beta {\cal H} (\{ w \},\{ \epsilon \})],
\label{PDF_full}
\end{equation}
where the hidden variables $\{x\}$ have been integrated out and $\beta$ is the inverse ``temperature" of the economic system. For a weak perturbation, Eq.~(\ref{PDF_full}) is expanded to the first order of $\{\epsilon\}$ as
\begin{equation}
P(\{ w \},\{ \epsilon \}) \simeq P(\{ w \}) \left(1 + \beta \sum_{\ell=1}^M \epsilon_\ell w_\ell \right) ,
\label{PDF_1st-order}
\end{equation}
where 
\begin{equation}
P(\{ w \}) \propto \exp [-\beta H(\{ w \})]
\label{PDF_0}
\end{equation}
is the PDF in the absence of $\{\epsilon\}$. Equation (\ref{PDF_1st-order}) enables us to calculate the induced change in $w_\ell$ by $\{\epsilon\}$ as 
\begin{equation}
\begin{split}
\langle w_\ell \rangle &= \int P(\{ w \},\{ \epsilon \}) w_\ell d{\{ w \}} \\
 &\simeq \beta \sum_{m=1}^M \epsilon_m \int P(\{ w \}) w_\ell w_m d{\{ w \}}\\
 &= \beta \sum_{m=1}^M \epsilon_m \langle w_\ell w_m \rangle_{0} ,
\end{split}
\label{eq:linear_response_cal}
\end{equation}
where $\langle \cdot \rangle_{0}$ denotes the ensemble average without perturbation. Comparison of Eqs. (\ref{eq:linear_response}) and (\ref{eq:linear_response_cal}) gives one of the outcomes of the FD theorem:
\begin{equation}
\bm{\chi} = \beta \bm{C}^{(0)},
\label{fdanzats}
\end{equation}
where 
$\bm{C}^{(0)}$ denotes a correlation matrix in the absence of external perturbations.
 
Making use of Eq.~(\ref{fdanzats}), we can rewrite the relation (\ref{wxw}) of economic ripple effects caused by the increase in shipments of final demand goods as
\begin{equation}
\begin{split}
\langle w_1\rangle &= C^{(0)}_{1m}\langle w_m \rangle, \\
\vdots \\
\langle w_M\rangle &= C^{(0)}_{Mm}\langle w_m \rangle. \\
\end{split}
\label{wcw}
\end{equation}
This is just one example of possible interindustrial relations derived from the present formulation. What we should emphasize here is that Eq.~(\ref{wcw}) has a rather general form in the framework of linear response.

For instance, we do not need to determine the temperature of the economic system for $\beta$. Even the assumption of Gibbs' ensemble, e.g., as given in Eqs. (\ref{PDF_full}) and (\ref{PDF_0}), may be too restrictive, because the assumption (\ref{PDF_1st-order}) about the PDF is sufficient to derive Eq.~(\ref{wcw}). We also recall Onsager's regression hypothesis~\cite{onsager1931a, onsager1931b}, on which the fluctuation-dissipation theorem relies. Once one accepts the hypothesis, one can readily derive Eq.~(\ref{wcw}). According to him, the response of a system in equilibrium to an external field shares an identical law with its response to a spontaneous fluctuation. In other words, the regression of spontaneous fluctuations at equilibrium takes place in the same way as the relaxation of non-equilibrium disturbances does. Let us suppose that the non-equilibrium disturbances $\langle w_i\rangle$ and $\langle w_m \rangle$ are linearly related through
\begin{equation}
\langle w_i\rangle = \kappa \langle w_m \rangle .
\label{onsager1}
\end{equation}
Accordingly, the spontaneous fluctuations $w_i$ and $w_m$ satisfy the same relation as Eq.~(\ref{onsager1}):
\begin{equation}
w_i = \kappa w_m .
\label{onsager2}
\end{equation}
The ensemble average of Eq.~(\ref{onsager2}) multiplied by $w_m$ on both hand sides determines the proportionality coefficient $\kappa$ as 
\begin{equation}
\kappa = C^{(0)}_{im} .
\label{onsager3}
\end{equation}
We thus see that Eq.~(\ref{wcw}) is directly derivable from Onsager's hypothesis.

We note that the correlation matrix appearing in Eqs.~(\ref{fdanzats}) and (\ref{wcw}) should be measured for a system not subject to any perturbations. However, the genuine correlation matrix $C^{\rm (G)}_{\ell m}$ determined by Eq.~(\ref{eq:C_true}) is possibly contaminated with various kinds of external economic shocks. While such forces may easily affect the stochastic motion of each $w$, it is legitimate to assume that their influence on the correlations among $w$'s are much weaker; otherwise, the external factors would have to work coherently to change ``springs" connecting pairs of $w$'s. This consideration justifies the replacement of $C^{(0)}_{\ell m}$ in Eq.~(\ref{wcw}) with $C^{\rm (G)}_{\ell m}$. 

\section{Interindustrial Relations}\label{sec:ic}

\figp{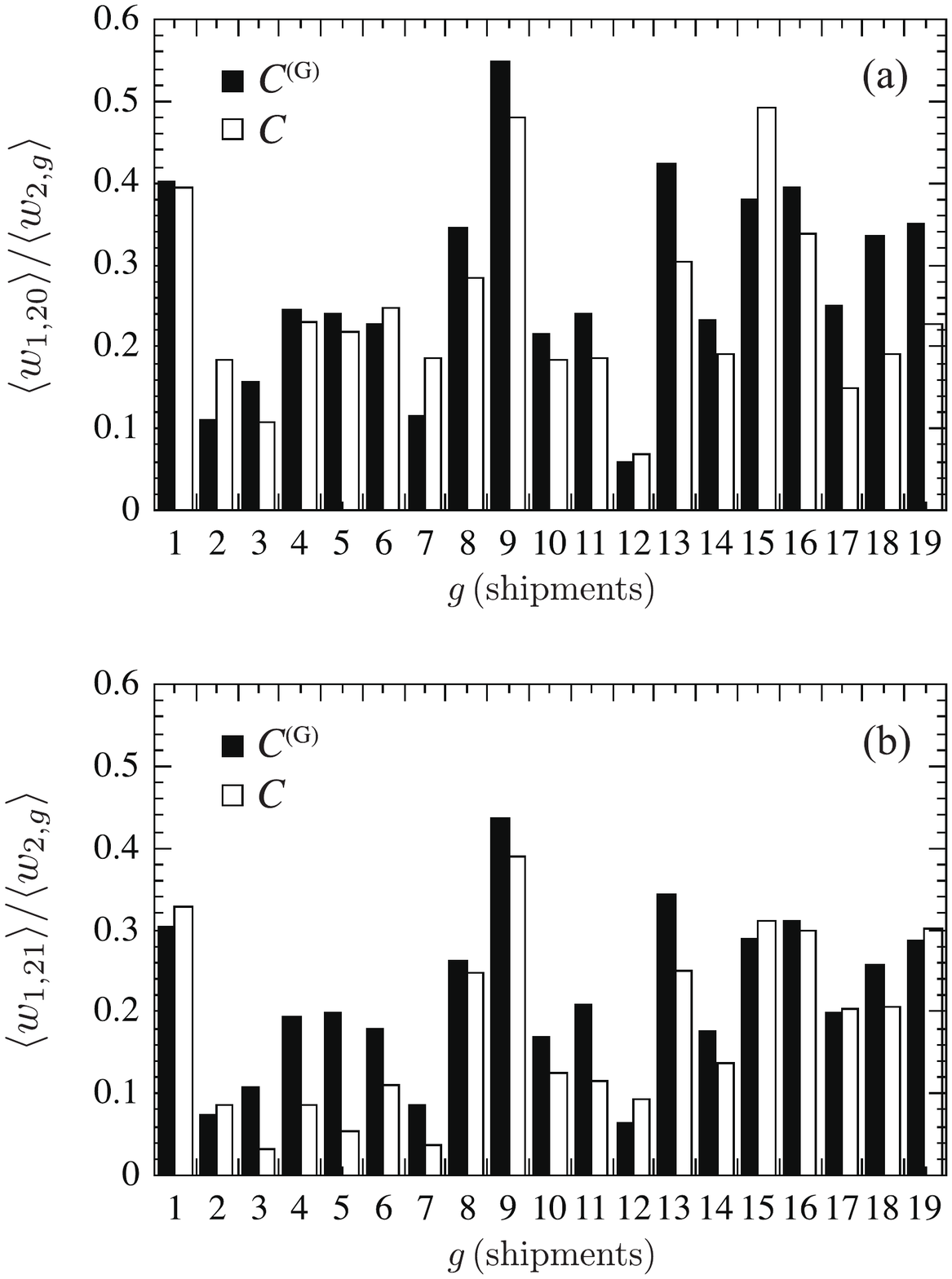}{Input-output interindustrial relations based on the genuine and original correlation matrices. The vertical axis in panel (a) indicates the extent to which the logarithmic growth rate of production of intermediate goods for Mining \& Manufacturing ($g=20$) is accordingly elevated when the logarithmic growth rate of shipments of each of the final demand goods specified on the horizontal axis is increased by one unit. Panel (b) is the same as panel (a), except that it shows the relationship between production of intermediate goods for Others ($g=21$) and shipments of each of the final demand goods.}{0.40}
{73.5981 131.503 543.63 765.531}

We are now in a position to quantitatively estimate the strength of the interindustrial relations by making use of the genuine correlation matrix through Eq.~(\ref{wcw}). In particular, we focus on ripple effects on production of intermediate goods that are triggered by applying an external stimulus to consumption of final demand goods: 
\begin{equation}
\langle w_{1,20}\rangle = C^{\rm (G)}_{1,20;2,g}\langle w_{2,g} \rangle, 
\label{eq:p20}
\end{equation}
\begin{equation}
\langle w_{1,21}\rangle = C^{\rm (G)}_{1,21;2,g}\langle w_{2,g} \rangle,
\label{eq:p21}
\end{equation}
with $g=1, 2, \cdots, 19$. 

The results for production of intermediate goods for Mining \& Manufacturing ($g=20$) are shown in the upper panel of Fig.~\ref{fig:production_g20and21}, in which those obtained with the original correlation matrix are also added for comparison. This figure shows an increase in the logarithmic growth rate of production of intermediate goods that is predicted from unit increment of the logarithmic growth rate of shipments of each of the final demand goods. As expected, increase in shipments of final demand goods with large weights, as represented by Manufacturing Equipment ($g=1$), Construction ($g=9$), Motor Vehicles ($g=15$), House Work ($g=16$), and Food \& Beverage ($g=19$), certainly causes large ripple effects on the production of intermediate goods. If the original correlation matrix is replaced with the genuine one, then the relative importance of species of final goods is interchanged between Construction and Motor Vehicles. This is understandable because sales of cars are sometimes promoted just for inventory adjustment, having no effect on the growth of production of intermediate goods. We also note that the original correlation matrix significantly underestimates the effects of Furniture \& Furnishing ($g=13$) and Nondurable Consumer Goods ($g=17,18,19$). It is noteworthy that Furniture \& Furnishing and Clothing \& Footwear, having much smaller weights than the major final demand goods, have comparable contributions; some feedback mechanism must be working through the inner loop in the economic system.

The lower panel in Fig.~\ref{fig:production_g20and21} shows the corresponding results for production of intermediate goods for Others ($g=21$), whose weight is one order of magnitude smaller than that of Mining \& Manufacturing. The important species of final demand goods are common in both categories of intermediate goods. In contrast, the original correlation matrix significantly underestimates the effects of different final demand goods such as those given by $g=3$ to $g=7$.

Presence of a correlation between two stochastic variables $A$ and $B$ does not indicate the existence of a mechanical connection between them. Actually, correlating $A$ and $B$ might be driven by a third variable $C$; then, there would be no causality relationship between $A$ and $B$. To address this question, we provide detailed information on phase relations in the business cycles identified in the previous study~\cite{bc}. Table~\ref{tab:phase60} lists phases of the cyclic motion of production, shipments, and inventory at $T=60$ and $40$ for each of the goods. We can see that shipments of final demand goods are ahead of or almost in phase with production of intermediate goods; the resolution limit (one month) is $6^{\circ}$ and $9^{\circ}$ for $T = 60$ and $40$, respectively. Electricity ($g=2$) and Communication \& Broadcasting ($g=3$) are exceptions to this observation. The wave of production arrives first and then, that of shipments follows for Electricity; the production activity for Communication \& Broadcasting behaves significantly out of phase with that averaged over goods. Since we have adopted a static approximation for the interindustrial relations, it may be more appropriate to average the phase relations over frequency. The results are shown in Fig.~\ref{fig:phaseav} and Table~\ref{tab:phaseav}. The frequency-averaged phase relations in the cyclic behavior of the economic fluctuations thus support our postulate that production of intermediate goods is driven by increasing shipments of final demand goods with a few exceptions. 

\begin{table}[htdp]
\caption{Phases of periodic oscillations with $T=60$ and $40$ of production (P), shipments (S), and inventory (I) for all goods, in the unit of degrees ranging from $-180$ to $180$. They are measured relative to the production of $g=20$.}
\begin{ruledtabular}
\begin{tabular}{ccccccc}
 & \multicolumn{3}{c}{$T=60$} & \multicolumn{3}{c}{$T=40$}\\ 
\textrm{Goods} & \multicolumn{1}{c}{\textrm{P}} & \multicolumn{1}{c}{\textrm{S}} & \multicolumn{1}{c}{\textrm{I}} & \multicolumn{1}{c}{\textrm{P}} & \multicolumn{1}{c}{\textrm{S}} & \multicolumn{1}{c}{\textrm{I}} \\
\hline
$1$&$-8.5$&$-8.2$&$-60.3$&$-9.8$&$-9.5$&$-117.7$\\
$2$&$-22.2$&$-45.3$&$-75.8$&$-35.4$&$-95.4$&$-130.4$\\
$3$&$-43.8$&$-40.9$&$-81.3$&$-92.2$&$-85.7$&$-133.6$\\
$4$&$-16.0$&$26.8$&$-85.6$&$-22.1$&$17.6$&$-135.9$\\
$5$&$4.4$&$61.6$&$-60.5$&$4.0$&$31.0$&$-117.8$\\
$6$&$2.5$&$15.8$&$-42.0$&$2.4$&$11.9$&$-88.2$\\
$7$&$-24.2$&$-7.0$&$-33.1$&$-40.3$&$-7.8$&$-65.1$\\
$8$&$-15.6$&$-4.3$&$-58.9$&$-21.3$&$-4.6$&$-116.1$\\
$9$&$13.8$&$37.0$&$-80.0$&$10.7$&$22.0$&$-132.9$\\
$10$&$56.2$&$28.4$&$-93.5$&$29.1$&$18.3$&$-139.6$\\
$11$&$9.2$&$83.8$&$-56.6$&$7.6$&$39.3$&$-113.4$\\
$12$&$-10.2$&$105.8$&$-54.3$&$-12.3$&$50.7$&$-110.5$\\
$13$&$7.8$&$48.1$&$-63.2$&$6.6$&$26.2$&$-120.6$\\
$14$&$-3.8$&$-4.2$&$-396.6$&$-4.0$&$-4.5$&$-74.6$\\
$15$&$1.2$&$-0.7$&$-24.2$&$1.1$&$-0.7$&$-40.3$\\
$16$&$9.7$&$27.2$&$-78.5$&$8.0$&$17.8$&$-132.1$\\
$17$&$11.6$&$35.4$&$-80.9$&$9.3$&$21.3$&$-133.4$\\
$18$&$-15.6$&$6.0$&$-66.1$&$-21.2$&$5.3$&$-123.2$\\
$19$&$44.6$&$55.2$&$-73.0$&$24.9$&$28.7$&$-128.6$\\
$20$&$0$&$15.8$&$-89.5$&$0$&$11.9$&$-137.8$\\
$21$&$7.4$&$40.7$&$-89.9$&$6.3$&$23.5$&$-138.0$\\
\hline
\textrm{Average} &$-0.6$&$25.2$&$-66.6$&$-0.6$&$16.9$&$-123.7$\\
\end{tabular}
\end{ruledtabular}
\label{tab:phase60}
\end{table}

\begin{table}[htdp]
\caption{Frequency-averaged phases of periodic motion of production (P), shipments (S), and inventory (I) for final demand and producer goods in the unit of degrees. The results for final demand goods were obtained by averaging over goods excluding $g=2$ and $3$; the numbers in parentheses are those obtained with all final demand goods.}
\begin{ruledtabular}
\begin{tabular}{cccc}
 & Final demand goods& \multicolumn{2}{c}{Producer goods}\\
 & & $g=20$ & $g=21$\\
\hline
P & $-0.69$ ($-2.35$) & $0$ & $0.99$ \\
S &  $2.99$ ($0.20$) & $5.84$ & $5.15$ \\
I & $-28.7$ ($-29.5$) & $-37.4$ & $-37.5$\\
\end{tabular}
\end{ruledtabular}
\label{tab:phaseav}
\end{table}

\figp{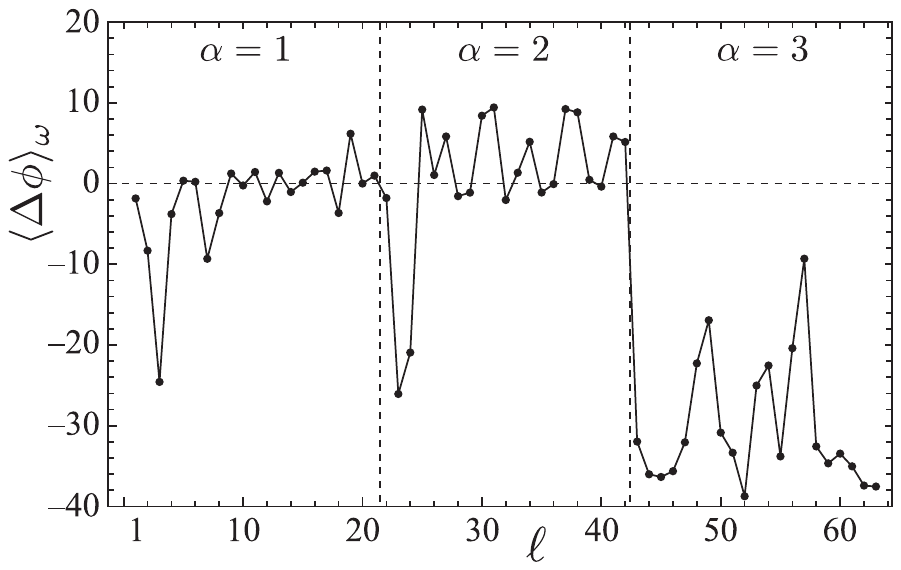}{Frequency-averaged phases of production, shipments, and inventory for each goods, measured relatively to production of $g=20$ in the unit of degrees.}{0.4}
{162.591 337.062 417.17 496.315}

\section{External Stimuli}\label{sec:iir}
Finally, we try to identify the presence of external stimuli hidden in real data by inversely using the linear response relationship (\ref{eq:linear_response}). The recent global economic crisis certainly has delivered an extremely large shock to the economic system of Japan, as is clearly shown in Fig.~\ref{fig:IIPaverage}. In our previous paper~\citep{bc}, however, we demonstrated that the crisis has simply increased the level of fluctuations associated with the dominant modes that were determined from the data during the normal time, instead of destroying the industrial structure itself; this is also manifested here, as shown in Fig.~\ref{fig:cjmv1}. And the information on collective movement of the IIP that we could extract from the dominant modes remains intact even in such an abnormal situation. This result thus conforms to the idea of Onsager's regression hypothesis, indicating the validity of the fluctuation-dissipation theory even in an economic system that is supposed to be far away from equilibrium.

Since approximation~(\ref{eq:C_true}) has been adopted for the correlation matrix, we consider only two independent external fields $\{ \eta_1, \eta_2 \}$ that are coupled to the normal coordinates $\{ a_1, a_2 \}$ associated with the two dominant eigenmodes $\bm{V}^{(1)}$ and $\bm{V}^{(2)}$, respectively. The total Hamiltonian (\ref{eq:total_H}) is therefore simplified to
\begin{equation}
{\cal H}=H(a_{1},a_{2}, \{x\}) - \eta_1 a_1 - \eta_2 a_2 .
\label{eq:total_H_reduced}
\end{equation}
The reduced external fields $\{ \eta \}$ in Eq.~(\ref{eq:total_H_reduced}) are derived from the original ones in Eq.~(\ref{eq:total_H}) through
\begin{equation}
\eta_{n}=\sum_{\ell=1}^{M}\epsilon_{\ell} V_{\ell}^{(n)} .
\end{equation}
Then, Eq.~(\ref{eq:linear_response}) is projected onto the two-dimensional reduced state space as
\begin{equation}
\left( {\begin{array}{@{\,}c@{\,}}
   {\left\langle {a_1 } \right\rangle }  \\
   {\left\langle {a_2 } \right\rangle }  \\
\end{array}} \right) = \left( {\begin{array}{@{\,}cc@{\,}}
   {\hat \chi _{11} } & {\hat \chi _{12} }  \\
   {\hat \chi _{21} } & {\hat \chi _{22} }  \\
\end{array}} \right)\left( {\begin{array}{@{\,}c@{\,}}
   {\eta _1 }  \\
   {\eta _2 }  \\
\end{array}} \right) ,
\label{eq:reduced_LRR}
\end{equation}
where
\begin{equation}
\left\langle {a_n } \right\rangle = \sum_{\ell=1}^{M} \left\langle {w_\ell } \right\rangle V_{\ell}^{(n)},
\label{eq:def_a_ind}
\end{equation}
and the reduced susceptibilities $\{ \hat \chi \}$ are defined as
\begin{equation}
\hat \chi _{mn}=\sum_{i=1}^{M}\sum_{j=1}^{M}V_{i}^{(m)}\chi_{ij}V_{j}^{(n)}.
\label{eq:reduced_kai}
\end{equation}
The relative values of $\{ \hat \chi \}$ with reference to ${\hat \chi_{11}}$ are calculated from $\bm{C}^{\rm (G)}$ as
\begin{equation}
\left( {\begin{array}{@{\,}cc@{\,}}
   {\hat \chi _{11} } & {\hat \chi _{12} }  \\
   {\hat \chi _{21} } & {\hat \chi _{22} }  \\
\end{array}} \right) = \beta
\left( {\begin{array}{@{\,}cc@{\,}}
   {1} & {1.30 \times 10^{-3}}  \\
   {1.30 \times 10^{-3}} & {0.433}  \\
\end{array}} \right).
\label{eq:reduced_kai_values}
\end{equation}
This results shows that the two eigenmodes are almost decoupled from each other, which is understandable from the orthogonality (\ref{lamaa}) of the normal coordinates.

One can obtain $\{ \eta \}$ using the inverse of Eq.~(\ref{eq:reduced_LRR}) along with Eqs.~(\ref{eq:def_a_ind}) and (\ref{eq:reduced_kai}), although it is not so straightforward. We first recall that $\langle w_\ell \rangle$ in the right-hand side of Eq.~(\ref{eq:def_a_ind}) is the deviation of $w_\ell$ from the equilibrium value induced by external perturbation, and not fluctuations of $w_\ell$ directly observed in the real data. We then identify $\langle w_\ell \rangle$ as residuals obtained by subtracting the long-period components arising from the inherent business cycles from moving average fluctuations of the IIP.

To extract $\langle w_\ell \rangle$, we first define the Fourier transform of the coefficients $a_n(t_j)$ as
\def\tn{N'}
\def\tw{\widetilde{w}}
\def\ta{\widetilde{a}}
\begin{equation}
a_n(t_j)=\frac{1}{\sqrt{\tn}}\sum_{k=1}^{\tn-1}\ta_n(\omega_k) \,e^{-i \omega_k t_j},
\label{fd2}
\end{equation}
with the Fourier frequency $\omega_k=2\pi k/(\tn\Delta t)$ and hence $\omega_k t_j =2\pi kj/\tn$.
The relevant long-period component $a^{\rm (LP)}_n(t_j)$ is obtained by limiting the sum over $k$ only to $k=1$ ($T=240$), $k=2$ ($T=120$), $k=4$ ($T=60$), and $k=6$ ($T=40$) or by summing all of the terms with periods larger than 2 years ($k\leqslant 9$) in Eq.~(\ref{fd2}). The formula for $\langle w_\ell \rangle$ is finally expanded as
\begin{equation}
\langle w_{\ell}(t) \rangle =\sum_{n=1}^{2} \overline{a_n(t)}\, V\ga^{(n)}-\sum_{n=1}^{2} a^{\rm (LP)}_n(t)\, V\ga^{(n)}.
\label{eq:d1_reduced}
\end{equation}

\figp{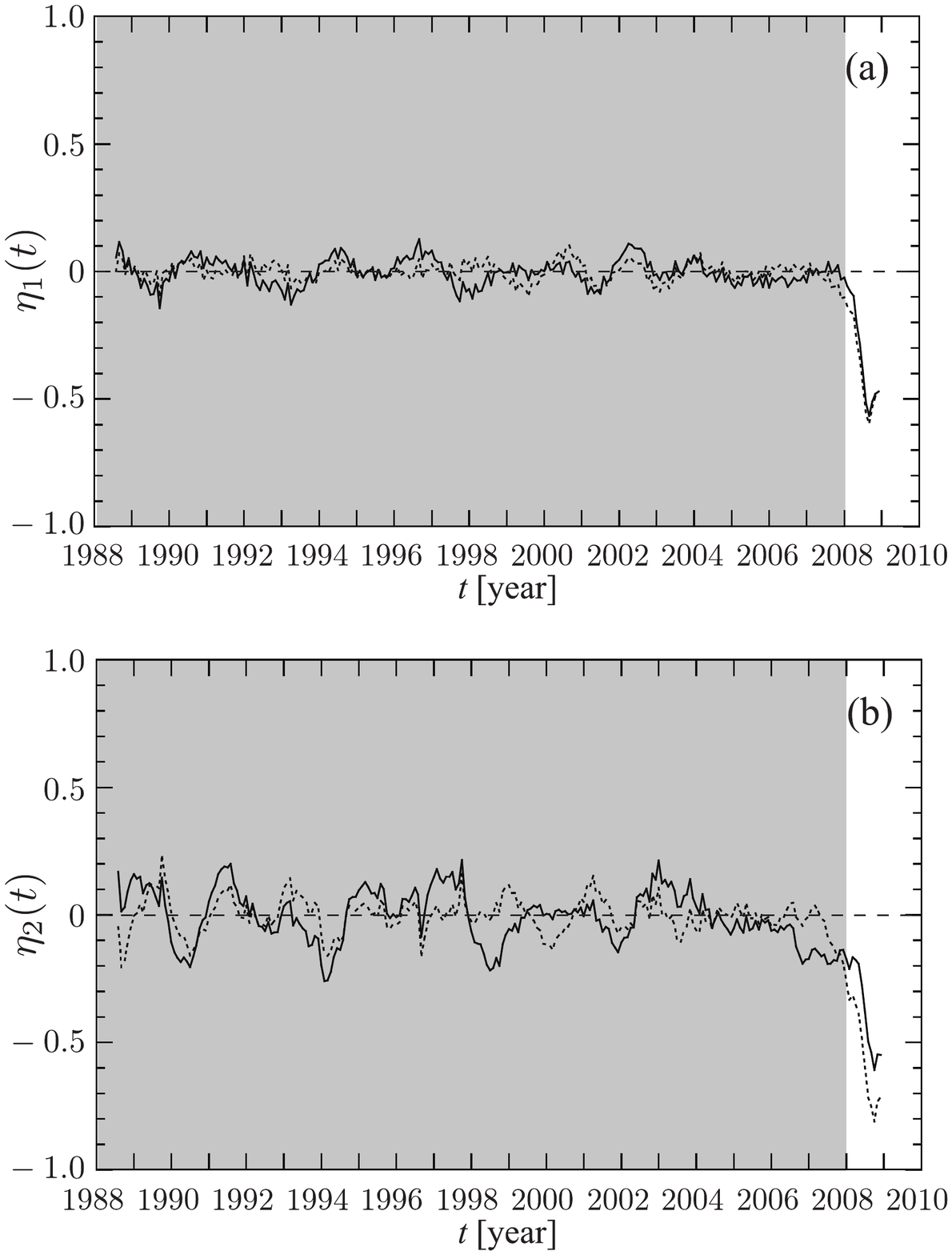}{External stimuli $\eta_{1}$ and $\eta_{2}$, derived from IIP data through the linear response relation (\ref{eq:reduced_LRR}), shown as a function of time in panels (a) and (b), respectively; the system is assumed to respond instantly to the applied external fields. The solid curves depict results obtained only with the terms of $k=1, 2, 4, \mathrm{and}\, 6$ in Eq.~(\ref{fd2}), and the dotted curves depict those calculated with the terms of $k\leqslant 9$. The gray shaded area is the same as depicted in Figs.~\ref{fig:IIPaverage} and \ref{fig:cjmv1}.}{0.4}
{45.4526 133.031 540.44 776.238}

Figure~\ref{fig:external_shocks}, for which we arbitrarily set $\beta=1$ in Eq.~(\ref{eq:reduced_kai_values}), shows the external fields $\eta_{1}$ and $\eta_{2}$ thus derived from $\langle w_\ell \rangle$. Two computational schemes were adopted to evaluate the long-period components in the IIP data, and no appreciable difference was observed between the two results. Here, the economic system was assumed to respond instantaneously to the applied external fields without any time delay. Referring to Fig.~\ref{fig:IIPaverage}, we clearly confirm that such a large external shock as manifested in $\eta_{1}$ causes the drastic drop in industrial activities in Japan. We also see that another large shock in $\eta_{2}$, which leads to reduction in inventory, immediately accompanies the first shock. In contrast, the maximum fluctuation levels of $\eta_{1}$ and $\eta_{2}$ are 0.1 and 0.2, respectively, in the normal period (before the end of 2007).

\section{Conclusion}\label{sec:concl}
This paper has described our attempt to utilize the fluctuation-dissipation theory for elucidating the nature of input-output correlations in the Japanese industry on the basis of IIP data. We were able to quantitatively estimate the strength of correlations between goods by using the genuine correlation matrix obtained in this study. We were also successful in extracting external stimuli over the last two decades. The noise reduction along with the RMT enabled us to detect economic signals hidden behind the complicated dynamics of the IIP. The strong coincidence between the sudden change in IIP data and the external shocks described here may prove that the present method is capable of predicting the input-output interindustrial relationship with a much higher time resolution than the annual resolution. We thus expect the results of this study to provide a new methodology for gaining deeper understanding of complex economic phenomena at a macroscopic level.

\begin{acknowledgments}
The present study was supported in part by {\it the Program for Promoting Methodological Innovation in Humanities and Social Sciences by Cross-Disciplinary Fusing} of the Japan Society for the Promotion of Science
and by the Ministry of Education, Science, Sports and Culture, Grants-in-Aid for Scientific Research (B), Nos. 20330060 (2008-10) and 22300080 (2010-12).
We would also like to thank Hiroshi Yoshikawa for providing continual advice and encouragement. 
\end{acknowledgments}

\end{document}